\documentclass[twocolumn,aps,prb,showpacs]{revtex4}
\usepackage{graphicx}
\usepackage{amsmath}
\usepackage{amsfonts}
\usepackage{amssymb}

\begin{document}

\author{Yaroslav Tserkovnyak and Arne Brataas}
\altaffiliation{Present address:
Department of Physics, Norwegian University of Science and Technology,
N-7491 Trondheim, Norway}
\affiliation{Lyman Laboratory of Physics, Harvard University, Cambridge,
Massachusetts 02138}
\author{Gerrit E. W. Bauer}
\affiliation{Department of NanoScience, Delft University of
Technology, 2628 CJ Delft, The Netherlands}
\title{Spin pumping and magnetization dynamics in metallic multilayers}

\begin{abstract}
We study the magnetization dynamics in thin ferromagnetic
films and small ferromagnetic particles in contact with paramagnetic
conductors. A moving magnetization vector 
causes \textquotedblleft pumping\textquotedblright\ of spins into adjacent
nonmagnetic layers. This spin transfer
affects the magnetization dynamics similar to the
Landau-Lifshitz-Gilbert phenomenology. The additional Gilbert damping is
significant for small ferromagnets, when the nonmagnetic layers efficiently
relax the injected spins, but the effect is reduced when a spin accumulation
build-up in the normal metal opposes the spin pumping.
The damping enhancement is governed by (and, in turn, can be used to
measure) the mixing conductance or spin-torque parameter of the
ferromagnet--normal-metal interface. Our theoretical findings
are confirmed by agreement with recent experiments in a
variety of multilayer systems.
\end{abstract}

\date{\today }
\maketitle


\section{Introduction}

Spin-polarized transport through magnetic multilayers is the physical origin
of many exciting phenomena such as giant magnetoresistance and
spin--current-induced magnetization reversal.\cite
{Gijs:ap97,Sloncz:mmm96,Sloncz:mmm99} It
has attracted attention in the basic physics community and industry over the
last decades, but there are still open fundamental questions. So far, the
main research activity has been focused on the dc transport properties of
these systems.

Ac magnetotransport has attracted considerably less attention than its dc
counterpart. In a recent paper,\cite{Tserkovnyak:prl02} we reported a novel
mechanism by which a precessing ferromagnet \textquotedblleft
pumps\textquotedblright\ a spin current into adjacent nonmagnetic conductors
proportional to the precession frequency, using a formalism analogous to
that for the adiabatic pumping of charges in mesoscopic systems.\cite
{Brouwer:prb98} We showed that spin pumping profoundly affects the dynamics
of nanoscale ferromagnets and thin films, by renormalizing fundamental
parameters such as the gyromagnetic ratio and Gilbert damping parameter,
in agreement with experiments.\cite{Mizukami:mmm01}

The switching characteristics of a magnetic system depends in an essential
way on the Gilbert damping constant $\alpha $. In magnetic field-induced
switching processes, for example, $\alpha $ governs the technologically
important magnetization reversal time of ferromagnetic particles. Its
typical intrinsic value\cite{Bhagat:prb74} $\alpha _{0}\lesssim 10^{-2}$ for
transition metal ferromagnets is smaller than its optimal value of $\alpha
\gtrsim 10^{-1}$ for the fastest switching rates.\cite{Kikuchi:jap56}
The present mechanism allows
engineering of the damping constant by adding passive nonmagnetic components
to the system and/or adjusting the geometry to control spin flow and
relaxation rates described in this paper, thus helping to create high-speed
magnetoelectronic devices. Also, in spin--current-induced magnetization
reversal, the critical switching current is proportional to $\alpha $.\cite
{Sloncz:mmm99}

For some time it has been understood that a ferromagnet--normal-metal
(\textit{F-N}) interface leads to a dynamical coupling between the
ferromagnetic magnetization and the spins of the conduction-band electrons
in the normal metal.\cite{Silsbee:prb79,Sloncz:mmm96,Sloncz:mmm99,
Berger:prb96,Brataas:prl00,Waintal:prb00}
More recently, several theoretical frameworks were put forward proposing a
mechanism for magnetization damping due to \textit{F-N} interfacial
processes.\cite{Berger:prb96,Wegrowe:prb00,Tserkovnyak:prl02} This
\textit{F-N} coupling becomes important in the limit of ultrathin
($\lesssim $10~nm) ferromagnetic films and can lead to a sizable
enhancement of the damping constant.

Our theory is based on a new physical picture, according to
which the ferromagnetic damping can be understood as
an adiabatic pumping of spins into the adjacent normal metals.\cite
{Tserkovnyak:prl02} This spin transfer is governed by the reflection and
transmission matrices of the system, analogous to the scattering theory
of transport and interlayer exchange
coupling. The microscopic expression for the enhanced Gilbert damping and
the renormalized gyromagnetic ratio can be calculated by simple models or by
first-principles band-structure calculations without adjustable parameters.
The present theory therefore allows quantitative predictions for the
magnetization damping in hybrid systems that can be tested by experiments.

The Gilbert damping constant in thin ferromagnetic films has been
experimentally studied\cite
{Heinrich:prl87,Platow:prb98,Mizukami:mmm01,Mizukami:mmm02,Urban:prl01} by
measuring ferromagnetic-resonance (FMR) linewidths. In the regime of
ultrathin ferromagnetic films, $\alpha $ was in some cases found to be quite
large in comparison with the bulk value $\alpha _{0}$, and sensitively
depending on the substrate and capping layer materials. For example, when a
20-\AA -thick permalloy (Py) film was sandwiched between two Pt layers, its
damping was found to be $\alpha \sim 10^{-1}$, but recovered its bulk value
$\alpha \sim 10^{-2}$ with a Cu buffer and cap.\cite{Mizukami:mmm01} Heinrich 
\textit{et al.} \cite{Heinrich:prl87} observed an enhanced damping of
$\lesssim$20~\AA -thick Fe films when they were grown on Ag bulk substrates
but no significant change in the damping constant was seen for films grown
on GaAs even when the film thickness was reduced down to several atomic
monolayers.\cite{Heinrich:priv} We will demonstrate here that our theory
explains all these experimental findings well.

Previously, we studied the situation when the normal-metal layers adjacent
to the ferromagnetic films are perfect spin sinks, so that the spin
accumulation in the normal metal vanishes.\cite{Tserkovnyak:prl02} Here this
theory is generalized to consider the spin accumulation which enables us
to explain experimental findings for various \textit{F-N} systems\cite
{Heinrich:prl87,Mizukami:mmm01,Mizukami:mmm02} in a unified framework based
on the spin-pumping picture.

The paper is organized as follow. In Sec.~\ref{pump} and the Appendix,
the basic formalism of the adiabatic spin-pump theory\cite{Tserkovnyak:prl02}
is derived using a scattering-matrix approach, and an alternative
derivation is given for finite
systems, which is based on the conservation of energy and angular momentum.
In Secs.~\ref{nfn} and \ref{fnn}, we solve the diffusion equation to
describe transport of injected spins in single and composite normal-metal
layers. The spin loss due to spin-orbit or other spin-flip processes is
accounted for, leading to an overall damping of the ferromagnetic
magnetization precession. In particular, we use the theory to analyze the
representative case of Gilbert damping in Py-Pt, Py-Cu, and Py-Cu-Pt
hybrids, showing an excellent agreement between our theory and the
experimental results.\cite{Mizukami:mmm01,Mizukami:mmm02} The last
Sec.~\ref{cd} is devoted to discussions and conclusions.

\section{Precession-induced spin pumping}

\label{pump}

Consider an \textit{N-F-N} junction as in Fig.~\ref{f1} and Fig.~\ref{f6}
in the Appendix. Without a voltage
bias, no spin or charge currents flow when the magnetization of the
ferromagnet is constant in time. When the magnetization direction starts
precessing (as, e.g., under the influence of an applied magnetic field), a
spin current $\mathbf{I}_{s}^{\text{pump}}$ is pumped
out of the ferromagnet.\cite{Tserkovnyak:prl02}
This current into a given \textit{N} layer depends on the complex-valued
parameter $A\equiv A_{r}+iA_{i}$ (the \textquotedblleft spin-pumping
conductance\textquotedblright ) by 
\begin{equation}
\mathbf{I}_{s}^{\text{pump}}=\frac{\hbar }{4\pi }\left( A_{r}\mathbf{m}
\times \frac{d\mathbf{m}}{dt}-A_{i}\frac{d\mathbf{m}}{dt}\right) \,.
\label{Is}
\end{equation}
Here, the time-dependent order parameter of the ferromagnet is a unit vector 
$\mathbf{m}(t)$, assuming a monodomain magnet with a spatially uniform
magnetization at all times.
A detailed derivation of Eq.~(\ref{Is}) based on
the scattering-matrix theory of transport is presented in the Appendix.
$A=g^{\uparrow \downarrow }-t^{\uparrow \downarrow }$ depends on the
scattering matrix of the ferromagnetic film since 
\begin{equation}
g^{\sigma \sigma ^{\prime }}\equiv \sum_{mn}\left[ \delta
_{mn}-r_{mn}^{\sigma }(r_{mn}^{\sigma ^{\prime }})^{\ast }\right]   \label{g}
\end{equation}
is the dimensionless dc conductance matrix\cite{Brataas:prl00,Waintal:prb00}
and 
\begin{equation}
t^{\uparrow \downarrow }\equiv \sum_{mn}t_{mn}^{\prime \uparrow
}(t_{mn}^{\prime \downarrow })^{\ast }\,. \label{t}
\end{equation}
Here (see Fig.~\ref{f6}), $r_{mn}^{\uparrow }$ ($r_{mn}^{\downarrow }$)
is a reflection
coefficient for spin-up (spin-down) electrons on the normal-metal side and
$t_{mn}^{\prime \uparrow }$ ($t_{mn}^{\prime \downarrow }$) is a transmission
coefficient for spin-up (spin-down) electrons across the ferromagnetic film
from the opposite reservoir into the normal-metal layer, where $m$ and $n$
label the transverse modes at the Fermi energy in the normal-metal films.
Note that the magnetization can take arbitrary directions; in particular,
$\mathbf{m}(t)$ may be far away from its equilibrium value. In such a case,
the scattering matrix itself can depend on the orientation of the
magnetization, and one has to use $A(\mathbf{m})$ in Eq.~(\ref{Is}).

When the ferromagnetic film is thicker than its transverse
spin-coherence length, $d>\pi
/(k^{\uparrow}_{\text{F}}-k^{\downarrow}_{\text{F}})$, where
$k^{\uparrow(\downarrow)}_{\text{F}}$
are the spin-dependent Fermi wave vectors, $t^{\uparrow
\downarrow }$ vanishes,\cite{Stiles:prb02} the spin pumping through a given 
\textit{F-N} interface is governed entirely by the interfacial mixing
conductance $A=g^{\uparrow \downarrow }\equiv g_{r}^{\uparrow \downarrow
}+ig_{i}^{\uparrow \downarrow }$, and we can consider only one of the
two interfaces. This is the regime we are focusing on in this paper.
Note that the conductance matrix $g^{\sigma \sigma ^{\prime }}$ defined
in Eq.~(\ref{g}) has to be
renormalized for highly transparent interfaces in columnar geometries
(by properly subtracting Sharvin resistance contributions from the
inverse conductance parameters), as discussed in Ref.~\onlinecite{Bauer:prep}.

\begin{figure}[pth]
\includegraphics[scale=0.35,clip=]{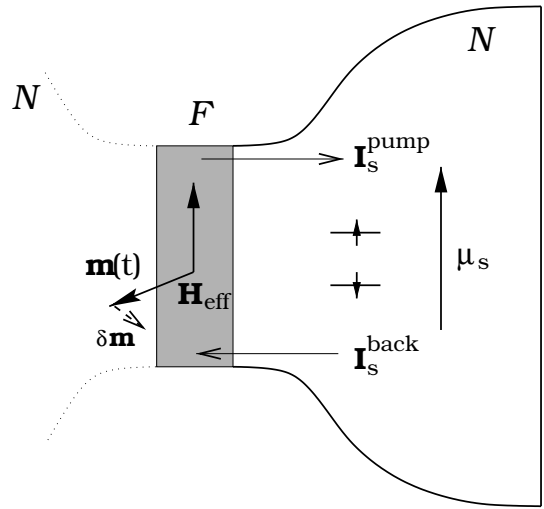}
\caption{A ferromagnetic film \textit{F} sandwiched between two
nonmagnetic reservoirs \textit{N}. For simplicity of the discussion in this
section, we mainly focus on the dynamics in one (right) reservoir while
suppressing the other (left), e.g., assuming it is insulating. The
spin-pumping current $\mathbf{I}_{s}^{\text{pump}}$ and the spin accumulation
$\text{\protect\boldmath$\protect\mu$}_{s}$ in the right
reservoir can be found by conservation of energy, angular momentum, and by
applying circuit theory to the steady state
$\mathbf{I} _{s}^{\text{pump}}=\mathbf{I}_{s}^{\text{back}}$.}\label{f1}
\end{figure}

As shown before,\cite{Tserkovnyak:prl02} the spin current [Eq.~(\ref{Is})]
leads to a damping of the ferromagnetic precession, resulting in a faster
alignment of the magnetization with the (effective) applied magnetic field
$\mathbf{H}_{\text{eff}}$. In the derivation by the time-dependent
scattering theory, 
the pumped spins are entirely absorbed by the attached ideal reservoirs.
In the following, it is shown that Eq.~(\ref{Is}) can be also derived for a
finite system by observing that the enhanced rate of damping is accompanied
by an energy flow out of the ferromagnet, until a steady state is established
in the combined \textit{F-N} system.
For simplicity, assume a magnetization which at time $t$ starts rotating
around the vector of the magnetic field,
$\mathbf{m}(t)\perp \mathbf{H}_{\text{eff}}$.
In a short interval of time $\delta t$, it slowly
(i.e., adiabatically) changes to
$\mathbf{m}(t+\delta t)=\mathbf{m}(t)+\delta\mathbf{m}$.
In the presence of a large but \textit{finite} nonmagnetic
reservoir without any spin-flip scattering attached to one side of the
ferromagnet, this process can be expected to induce a (small) nonvanishing
spin accumulation 
\begin{equation}
\text{\boldmath$\mu $}_{s}\equiv \int d\epsilon \mbox{Tr}[\text{\boldmath$
\hat{\sigma}$}\hat{f}(\epsilon )]\,,  \label{SA}
\end{equation}
where $\text{\boldmath$\hat{\sigma}$}$ is the Pauli matrix vector and $\hat{f
}(\epsilon )$ is the $2\times 2$ matrix distribution function at a given
energy $\epsilon$ in the reservoir.\cite{Brataas:prl00}
For a slow enough variation of $
\mathbf{m}(t)$, this nonequilibrium spin imbalance must flow back into the
ferromagnet, canceling any spin current generated by the magnetization
rotation, since, due to the adiabatic assumption, the system is always in
a steady state.

Let us assume for the moment that the spins are accumulated in the reservoir
along the magnetic field,
$\text{\boldmath$\mu$}_{s}\parallel\mathbf{H}_{\text{eff}}$. Flow
of $N_s$ spins into the normal metal transfers
energy $\Delta E_N=N_s\mu_{s}/2$ and angular-momentum
$\Delta L_N=N_s\hbar/2$ (directed along
$\mathbf{H}_{\text{eff}}$).
By the conservation laws,
$\Delta E_F=-\Delta E_N$ and $\Delta L_F=-\Delta L_N$, for the corresponding
values in the ferromagnet. Using the magnetic energy
$\Delta E_F=\gamma\Delta L_FH_{\text{eff}}$,
where $\gamma$ is the
absolute gyromagnetic ratio of the ferromagnet, we then find that
$N_s\mu_{s}/2=\gamma N_s(\hbar/2)H_{\text{eff}}$. It then follows that
$\mu_{s}=\hbar\gamma H_{\text{eff}}=\hbar\omega$, where $\omega=\gamma H_{
\text{eff}}$ is the Larmor frequency of precession in the effective
field: The spin-up and spin-down chemical potentials in the normal metal are
split by $\mu_{s}=\hbar\omega$, the energy corresponding to the frequency
of the perturbation. For a finite angle $\theta $ between $\text{\boldmath$
\mu $}_{s}$ and $\mathbf{H}_{\text{eff}}$, the same reasoning would lead to $
\mu _{s}=\hbar \omega \cos \theta $, which is smaller than
the \textquotedblleft energy boost\textquotedblright\ $\hbar \omega $ of the
time-dependent perturbation, thus justifying our initial guess.

We can employ magnetoelectronic circuit theory\cite {Brataas:prl00} to derive
an expression for the backflow of spin current
$\mathbf{I}_{s}^{\text{back}}$ which, as argued above, has to be equal to the
pumping current $\mathbf{I}_{s}^{\text{pump}}=\mathbf{I}_{s}^{\text{back}}$: 
\begin{align}
\mathbf{I}_{s}^{\text{back}}& =\frac{1}{2\pi }\left( g_{r}^{\uparrow
\downarrow }\text{\boldmath$\mu $ }_{s}+g_{i}^{\uparrow \downarrow }
\mathbf{m}\times \text{\boldmath$\mu$}_{s}\right)   \notag \\
& =\frac{\hbar }{4\pi }\left( g_{r}^{\uparrow \downarrow }\mathbf{m}\times 
\frac{d\mathbf{m}}{dt}-g_{i}^{\uparrow \downarrow }\frac{d\mathbf{m}}{dt}
\right) \,.
\end{align}
Here, we used $\mu _{s}=\hbar \omega $ and $\text{\boldmath$\mu $}_{s}\perp 
\mathbf{m}$, since by the conservation of angular momentum, the spin
transfer is proportional to the change in the direction $\delta \mathbf{m}
\perp \mathbf{m}$. We thus recover Eq.~(\ref{Is}) for the case of a single
and finite reservoir. It is easy to repeat the proof for an arbitrary
initial alignment of $\mathbf{m}(t)$ with $\mathbf{H}_{\text{eff}}$.
Furthermore, a straightforward generalization of this discussion to the case
of the \textit{N-F-N} sandwich structure recovers our previous result
[Eq.~(\ref{Is})].

The expressions for the adiabatic spin pumping are not the whole story,
since spin-flip scattering is an important fact of life in
magnetoelectronics. In Ref.~\onlinecite{Tserkovnyak:prl02}, we only
considered the extreme situation where the normal-metal layer is a perfect
spin sink, so that all spins injected by $\mathbf{I}_{s}^{\text{pump}}$
relax by spin-flip processes or leave the system; the total spin current
through the contact was, therefore, approximated by
$\mathbf{I}_{s}\approx\mathbf{I}_{s}^{\text{pump}}$ and
$\mathbf{I}_{s}^{\text{back}}\approx0$. Here, we
generalize that treatment to self-consistently take into account the spin
build-up in the normal metal at dynamic
equilibrium. We then find the contribution to $\mathbf{I}_{s}$ due to the
spin--accumulation-driven current $\mathbf{I}_{s}^{\text{back}}$ back into
the ferromagnet:
\begin{equation}
\mathbf{I}_{s}=\mathbf{I}_{s}^{\text{pump}}-\mathbf{I}_{s}^{\text{back} }\,,
\label{sc}
\end{equation}
which vanishes in the absence of spin-flip scattering.

The spin current out of the ferromagnet carries angular momentum
perpendicular to the magnetization direction. By conservation of angular
momentum, the spins ejected by $\mathbf{I}_{s}$ correspond to a
torque $\text{\boldmath$\tau$}=-\mathbf{I}_{s}$ on the ferromagnet. If
possible interfacial spin-flip processes are disregarded, the torque $\text{
\boldmath$\tau$}$ is entirely transferred to the coherent magnetization
precession. The dynamics of the ferromagnet can then be described by a
generalized Landau-Lifshitz-Gilbert (LLG) equation\cite
{Gilbert:pr55,Sloncz:mmm96} 
\begin{equation}
\frac{d\mathbf{m}}{dt}=-\gamma\mathbf{m}\times\mathbf{H}_{\text{eff}}
+\alpha_{0}\mathbf{m}\times\frac{d\mathbf{m}}{dt}+\frac{\gamma}{M_{s}V}
\mathbf{I}_{s}\,,  \label{llg}
\end{equation}
where $\alpha_{0}$ is the dimensionless bulk Gilbert damping constant,
$M_{s}$ is the saturation magnetization of the ferromagnet,
and $V$ is its volume. The intrinsic bulk
constant $\alpha_{0}$ is smaller than the total Gilbert damping
$\alpha=\alpha_{0}+\alpha^{\prime}$. The additional damping $
\alpha^{\prime}$ caused by the spin pumping is observable in, for example,
FMR spectra and is the main object of interest here.

\section{Spin--accumulation-driven backflow in the \textit{F-N} and
\textit{N-F-N} multilayers}
\label{nfn}

\begin{figure}[pth]\includegraphics[scale=0.4,clip=]{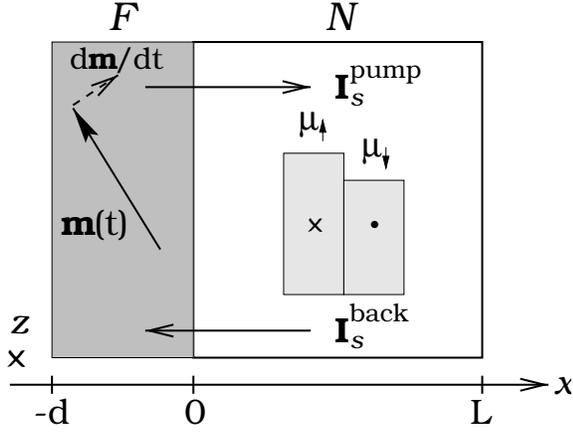}
\caption{Schematic view of the \textit{F-N} bilayer. Precession of the
magnetization direction $\mathbf{m}(t)$ of the ferromagnet \textit{F} pumps
spins into the adjacent normal-metal layer \textit{N} by inducing a spin
current $\mathbf{I}_{s}^{\text{pump}}$. This leads to a build-up of the
normal-metal spin accumulation which either relaxes by spin-flip
scattering or flows back into the ferromagnet as
$\mathbf{I}_{s}^{\text{back}}$. In contrast to Fig.~\protect\ref{f1}, the 
\textit{N} layer here is not an ideal reservoir but rather a film of the
same cross section as the magnetic layer \textit{F}; the spin accumulation
is position ($x$) dependent.}
\label{f2}
\end{figure}

The precession of the magnetization does not cause any charge current in
the system. The spin accumulation or nonequilibrium chemical potential
imbalance $\text{\boldmath$\mu $}_{s}(x)$ [similar to Eq. (\ref{SA}), but
spatially dependent now] in the normal metal is a vector, which depends on
the distance from the interface $x$, $0<x<L$, where $L$ is the thickness of
the normal-metal film, see Fig.~\ref{f2}.

When the ferromagnetic magnetization steadily rotates around the $z$ axis,
$\mathbf{m}\times \mathbf{\dot{m}}$ and the normal-metal spin accumulation
$\text{\boldmath$\mu $}_{s}(x)$ are
oriented along $z$, as depicted in Fig.~\ref{f2}. There is no spin
imbalance in the ferromagnet, because $\text{\boldmath$\mu $}_{s}$ is
perpendicular to the magnetization direction $\mathbf{m}$. As shown below,
the time-dependent $\text{\boldmath$\mu $}_{s}$ is also perpendicular
to $\mathbf{m}$ even in the case of a \textit{precessing} ferromagnet
with time-dependent instantaneous rotation axis, as long as the precession
frequency $\omega $ is smaller than the spin-flip rate $\tau _{\text{sf}
}^{-1}$ in the normal metal.

The spin accumulation diffuses into the normal metal as 
\begin{equation}
i\omega \text{\boldmath$\mu $}_{s}=D\partial _{x}^{2}\text{\boldmath$\mu $}
_{s}-\tau _{\text{sf}}^{-1}\text{\boldmath$\mu $}_{s}\,,  \label{de}
\end{equation}
where $D$ is the diffusion coefficient. The boundary conditions are
determined by the continuity of the spin current from the ferromagnet into
the normal metal at $x=0$ and the vanishing of the spin current at the outer
boundary $x=L$:
\begin{align}
x& =0:~\partial _{x}\text{\boldmath$\mu $}_{s}=-2(\hbar \mathcal{N}SD)^{-1}
\mathbf{I}_{s}\,,  \notag \\
x& =L:~\partial _{x}\text{\boldmath$\mu $}_{s}=0\,,  \label{bc}
\end{align}
where $\mathcal{N}$ is the (one-spin) density of states in the film and $S$
is the area of the interface. The solution to Eq.~(\ref{de}) with the
boundary conditions [Eqs.~(\ref{bc})] is 
\begin{equation}
\text{\boldmath$\mu $}_{s}(x)=\frac{\cosh \kappa (x-L)}{\sinh \kappa L}\frac{
2\mathbf{I}_{s}}{\hbar \mathcal{N}SD\kappa }  \label{sn}
\end{equation}
with the wave vector $\kappa =\lambda _{\text{sd}}^{-1}\sqrt{1+i\omega \tau
_{\text{sf}}}$, where $\lambda _{\text{sd}}\equiv \sqrt{D\tau _{\text{sf}}}$
is the spin-flip diffusion length in the normal metal.
In Ref.~\onlinecite{Brataas:prb02} we used arguments similar to those in the
present paper to calculate the spin accumulation (\ref{sn}) generated by the
precessing magnetization. While the size of the effect and its relevance for
spintronic applications are detailed in Ref.~\onlinecite{Brataas:prb02}, in
this work we focus on the role of the spin accumulation in the
dynamics of the ferromagnetic magnetization.

We assume in the following that the precession frequency $\omega$ is smaller
than the spin-flip relaxation rate $\omega \ll \tau _{\text{sf}}^{-1}$ so
that $\kappa\approx\lambda_{\text{sd}}^{-1}$. For a static applied field of
1~T, typically $\omega \sim 10^{11}$~s$^{-1}$. The elastic
scattering rate corresponding to a mean free path of
$\lambda_{\text{el}}\sim10$~nm is $
\tau _{\text{el}}^{-1}\sim 10^{14}$~s$^{-1}$. Consequently, the derivation
below is restricted to metals with a ratio of spin-conserved to spin-flip 
scattering times $\epsilon \equiv \tau _{\text{el}}/\tau _{\text{sf}}\gtrsim
10^{-3}$. In practice,\cite{Meservey:prl78} this condition is easily satisfied
with higher impurity atomic numbers $Z$ (as $\epsilon$ scales
as\cite{Abrikosov:zetf62} $Z^{4}$). The high-frequency limit
$\omega\gtrsim\tau_{sf}^{-1}$, on the other hand, is relevant for
hybrids with little spin-flip scattering in the normal metal, and was
discussed in the context of the spin-battery concept.\cite{Brataas:prb02}
Nevertheless, we will see that a sizable Gilbert damping enhancement
requires a large spin-flip probability $\epsilon \gtrsim10^{-1}$
(thereby guaranteeing that $\omega\ll\tau_{sf}^{-1}$) unless the frequency
is comparable with the elastic scattering rate in the normal metal.
The latter regime will not be treated in this paper.

Using relation $D=v_{\text{F}}^{2}\tau_{\text{el}}/3$ between the diffusion
coefficient $D$ (in three dimensions), the Fermi velocity $v_{\text{F}}$, and
the elastic scattering time $\tau_{\text{el}}$, we find for the spin-diffusion
length
\begin{equation}
\lambda_{\text{sd}}=v_{\text{F}}\sqrt{\tau_{\text{el}}\tau_{\text{sf}}
/3}\,.\label{sd}
\end{equation}
An effective energy-level spacing of the states participating in the spin-flip
scattering events in a thick film can be defined by
\begin{equation}
\delta_{\text{sd}}\equiv(\mathcal{N}S\lambda
_{\text{sd}})^{-1}\,.\label{td}
\end{equation}
The spin--accumulation-driven spin current $\mathbf{I}_{s}^{\text{back}}$
through the interface reads\cite{Brataas:epjb01} 
\begin{align}
\mathbf{I}_{s}^{\text{back}}& =\frac{1}{8\pi }\left[ 2g
_{r}^{\uparrow \downarrow }\text{\boldmath$\mu $}_{s}(x=0)+2g
_{i}^{\uparrow \downarrow }\mathbf{m}\times \text{\boldmath$\mu $}
_{s}(x=0)\right.   \notag \\
& +\left. \left(g^{\uparrow \uparrow }+g^{\downarrow
\downarrow }-2g_{r}^{\uparrow \downarrow }\right) \left( \mathbf{m}
\cdot \text{\boldmath$\mu $}_{s}(x=0)\right) \mathbf{m}\right] \,.
\label{ne}
\end{align}
Substituting Eq.~(\ref{sn}) into Eq.~(\ref{ne}), we
find for the total spin current [Eq.~(\ref{sc})] 
\begin{align}
\mathbf{I}_{s}& =\mathbf{I}_{s}^{\text{pump}}-\frac{\beta }{2}\left[ 2
g_{r}^{\uparrow \downarrow }\mathbf{I}_{s}+2g_{i}^{\uparrow
\downarrow }\mathbf{m}\times \mathbf{I}_{s}\right.   \notag \\
& \left. +\left(g^{\uparrow \uparrow }+g^{\downarrow
\downarrow }-2g_{r}^{\uparrow \downarrow }\right) \left( \mathbf{m}
\cdot \mathbf{I}_{s}\right) \mathbf{m}\right] \,,  \label{sc1}
\end{align}
where the spin current returning into the ferromagnet is governed by the
\textquotedblleft backflow\textquotedblright\ factor\ $\beta $, 
\begin{equation}
\beta \equiv \frac{\tau _{\text{sf}}\delta_{\text{sd}}/h}{\tanh (L/\lambda _{
\text{sd}})}\,.  \label{beta}
\end{equation}
When the normal metal is shorter than the spin-diffusion length ($L\ll
\lambda _{\text{sd}}$), $\beta \rightarrow \tau _{\text{sf}}\delta /h$,
where $\delta =(\mathcal{N}SL)^{-1}$ is the energy-level splitting. In the
opposite regime of thick normal metals ($L\gg \lambda _{\text{sd}}$), $\beta
\rightarrow \tau _{\text{sf}}\delta _{\text{sd}}/h$. Basically, $\beta $
[Eq.~(\ref{beta})] is therefore the ratio between the energy level spacing
of the normal-metal film with a thickness $L_{\text{sf}}=\min (L,\lambda _{
\text{sd}})$ and the spin-flip rate.

By inverting Eq.~(\ref{sc1}), we may express the total spin current $\mathbf{
I}_{s}$ in terms of the pumped spin current $\mathbf{I}_{s}^{\text{pump}}$
[Eq.~(\ref{Is})] 
\begin{align}
\mathbf{I}_{s}& =\left[ 1+\beta g_{r}^{\uparrow \downarrow }+\frac{
(\beta g_{i}^{\uparrow \downarrow })^{2}}{1+\beta g
_{r}^{\uparrow \downarrow }}\right] ^{-1}  \notag \\
& \left( 1-\frac{\beta g_{i}^{\uparrow \downarrow }}{1+\beta g
_{r}^{\uparrow \downarrow }}\mathbf{m}\times \right) \mathbf{I}_{s}^{\text{
pump}}\,.  \label{sc2}
\end{align}
After substituting Eq.~(\ref{Is}) into Eq.~(\ref{sc2}), we recover the form
of Eq.~(\ref{Is}) for the total spin current $\mathbf{I}_{s},$ but with a
redefined spin-pumping conductance $\tilde{A}\equiv \tilde{A}_{r}+i\tilde{
A}_{i}$ 
\begin{equation}
\mathbf{I}_{s}=\frac{\hbar }{4\pi }\left( \tilde{A}_{r}\mathbf{m}\times 
\frac{d\mathbf{m}}{dt}-\tilde{A}_{i}\frac{d\mathbf{m}}{dt}\right) \,.
\label{Iss}
\end{equation}
$\tilde{A}$ can be expressed in terms of the mixing conductance $
g^{\uparrow \downarrow }$ and the
backflow factor $\beta $ by 
\begin{align}
\left( 
\begin{array}{c}
\tilde{A}_{r} \\ 
\tilde{A}_{i}
\end{array}
\right) & =\left( 
\begin{array}{cc}
1 & \beta g_{i}^{\uparrow \downarrow }(1+\beta g
_{r}^{\uparrow \downarrow })^{-1} \\ 
-\beta g_{i}^{\uparrow \downarrow }(1+\beta g_{r}^{\uparrow
\downarrow })^{-1} & 1
\end{array}
\right)   \notag \\
& \left[ 1+\beta g_{r}^{\uparrow \downarrow }+\frac{(\beta g
_{i}^{\uparrow \downarrow })^{2}}{1+\beta g_{r}^{\uparrow \downarrow
}}\right] ^{-1}\left( 
\begin{array}{c}
g_{r}^{\uparrow \downarrow } \\ 
g_{i}^{\uparrow \downarrow }
\end{array}
\right) \,.  \label{ps}
\end{align}
It has been shown\cite{Xia:prb02} that for realistic \textit{F-N}
interfaces $g_{i}^{\uparrow \downarrow }\ll g_{r}^{\uparrow \downarrow }$,
so that $g^{\uparrow \downarrow }\approx g_{r}^{\uparrow \downarrow }$.
(The latter approximation will be implied for the rest of the paper.)
In this important regime, $\tilde{A}_{i}$
vanishes and the term proportional to $\tilde{A}_{r}$ in Eq.~(\ref{Iss}) has
the same form as and therefore enhances the phenomenological Gilbert
damping. This can be easily seen after substituting Eq.~(\ref{Iss}) into
Eq.~(\ref{llg}): The last term on the right-hand side of Eq.~(\ref{llg}) can
be combined with the second term by defining the total Gilbert damping
coefficient $\alpha =\alpha _{0}+\alpha ^{\prime }$, where 
\begin{equation}
\alpha ^{\prime }=\left[ 1+g^{\uparrow \downarrow }\frac{\tau _{
\text{sf}}\delta_{\text{sd}}/h}{\tanh (L/\lambda _{\text{sd}})}\right] ^{-1}
\frac{g_{L}g^{\uparrow \downarrow }}{4\pi\mu}  \label{aa}
\end{equation}
is the additional damping constant due to the interfacial \textit{F-N}
coupling. Here, $g_{L}$ is the $g$ factor and $\mu$ is the total film
magnetic moment in units of $\mu _{\text{B}}$.
Equation.~(\ref{aa}) is the main result of this section.
When $L\rightarrow\infty$,
Eq.~(\ref{aa}) reduces to a simple result:
$\alpha ^{\prime }=g_{L}g_{\text{eff}}^{\uparrow\downarrow}/(4\pi\mu)$,
where
\begin{equation}
\frac{1}{g_{\text{eff}}^{\uparrow\downarrow}}=\frac{1}{g^{\uparrow\downarrow}}
+R_{\text{sd}}\,.
\label{ser}
\end{equation}
Here $R_{\text{sd}}=\tau_{\text{sf}}\delta_{\text{sd}}/h$ is the resistance
(per spin, in units of $h/e^2$) of the normal-metal layer of thickness
$\lambda_{\text{sd}}$. [Which follows from the Einstein's relation
$\sigma=e^2D\mathcal{N}$ connecting conductivity $\sigma$ with the
diffusion coefficient $D$, and using Eq.~(\ref{td}).]
It follows that the effective spin pumping out of the ferromagnet is
governed by $g_{\text{eff}}^{\uparrow\downarrow}$, i.e.,
the conductance of the \textit{F-N}
interface in series with diffusive normal-metal film with thickness
$\lambda_{\text{sd}}$.\cite{Bauer:prep}

The prefactor on the right-hand side of Eq.~(\ref{aa}) suppresses the
additional Gilbert damping\cite{Tserkovnyak:prl02} due to the spin angular
momentum that diffuses back into the ferromagnet. It was disregarded
in Ref.~\onlinecite{Tserkovnyak:prl02} where the normal metal was viewed
as a perfect spin sink. Because spins accumulate in the normal metal
perpendicular to the ferromagnetic magnetization, the
spin--accumulation-driven transport across the \textit{F-N} contact, as well
as the spin pumping, is governed by a mixing conductance. This explains why
the other components of the conductance matrix [Eq.~(\ref{g})] do not enter
Eq.~(\ref{aa}).

We now estimate the numerical values of the parameters in Eq.~(\ref{aa}) for
transition metal ferromagnets Fe, Co, and Ni, in contact with relatively
clean simple normal metals Al, Cr, Cu, Pd, Ag, Ta, Pt, and Au. For an
isotropic electron gas, $\mathcal{N}=k_{\text{F}}^{2}/(\pi hv_{
\text{F}})$. Using Eqs.~(\ref{sd}) and (\ref{td}), we find
$h/(\delta_{\text{sd}}\tau _{\text{sf}})=4\sqrt{\epsilon/3}N_{\text{ch}}$,
where $N_{\text{ch}}=Sk_{\text{F}}^{2}/(4\pi)$ is the number of transverse
channels in the normal metal and $\epsilon\equiv\tau _{\text{el}}/\tau
_{\text{sf}}$ is the spin-flip probability at each scattering.
In Ref.~\onlinecite{Xia:prb02}, $g^{\uparrow \downarrow }$
was calculated for
Co-Cu and Fe-Cr interfaces by first-principles band-structure calculations.
It was found that irrespective of the interfacial disorder, 
$g^{\uparrow \downarrow }\approx N_{\text{ch}}$
for these material combinations. As shown in
Ref.~\onlinecite{Bauer:prep}, $g^{\uparrow\downarrow}$ has to be
renormalized in such limit, making the effective
conductances about twice as large.
We thus arrive at an estimate 
\begin{equation}
\frac{\alpha _{\infty }^{\prime }}{\alpha ^{\prime }}\approx 1+\left[ \sqrt{
\epsilon }\tanh (L/\lambda _{\text{sd}})\right] ^{-1}\,,  \label{sp11}
\end{equation}
where $\alpha _{\infty }^{\prime }=g_{L}g^{\uparrow \downarrow }/(4\pi\mu)$
is the Gilbert damping enhancement assuming infinite spin-flip rate in the
normal metal $\tau _{\text{sf}}\rightarrow 0$, i.e., treating it as
a perfect spin sink.\cite{Tserkovnyak:prl02}

It follows that only for a high spin-flip probability $\epsilon \gtrsim
10^{-2}$, the normal-metal film can be a good spin sink so that $\alpha
^{\prime }\sim \alpha _{\infty }^{\prime }$. This makes the lighter metals,
such as Al, Cr, and Cu, as well as heavier metals with only
\textit{s} electrons in the conduction band,
such as Ag, Au, and Ta less effective spin
sinks since these metals have a relatively small spin-orbit coupling,
typically corresponding to $\epsilon \lesssim 10^{-2}$ .\cite
{Meservey:prl78,Bergmann:zpb82,Yang:prl94} Heavier elements
with $Z\gtrsim 50$ and $p$ or $d$ electrons in the conduction band,
such as Pd and Pt, on the other hand,
can be good or nearly perfect spin sinks as they
have a much larger $\epsilon \gtrsim 10^{-1}$.\cite{Meservey:prl78} This
conclusion explains the hierarchy of the observed Gilbert damping
enhancement in Ref.~\onlinecite{Mizukami:mmm01}: Pt has about 2
electrons per atom in the conduction band, which are hybridized with
\textit{d} orbitals, and a large atomic number $Z=78$
and, consequently, leads to a large magnetization damping enhancement in the 
\textit{N-F-N} sandwich for thin ferromagnetic films. Pd which is above Pt
in the periodic table having similar atomic configuration but smaller atomic
number $Z=46$ leads to a sizable damping, but smaller than for Pt by a
factor of 2. Ta is a heavy element, $Z=73$, but has only \textit{s} electrons
and the damping enhancement is an order of magnitude smaller than
in Pt. Finally, Cu is a relatively light element, $Z=29$, with
\textit{s} electrons only and does not cause an observable damping
enhancement at all.
According to Eq.~(\ref{sp11}), a sufficiently thick active layer, $L\gtrsim
\lambda _{\text{sd}}$, is also required for a sizable spin relaxation.

The limit of a large ratio of spin-flip to non-spin-flip scattering $
\epsilon\sim1$ deserves special attention. In this regime, Eq.~(\ref{sp11})
does not hold, since by using the diffusion equation [Eq.~(\ref{de})] and
boundary conditions [Eqs.~(\ref{bc})] we implicitly assumed that $\epsilon\ll
1$. If $\epsilon\gtrsim10^{-1}$, on the other hand, even interfacial
scattering alone can efficiently relax the spin imbalance, and such
films, therefore, are good or nearly perfect spin sinks (so that $
\alpha^{\prime}\sim\alpha_{\infty}^{\prime}$), regardless of their thickness
(in particular, they can be thinner than the elastic mean free path).

\begin{figure}[pth]
\includegraphics[scale=0.45,clip=]{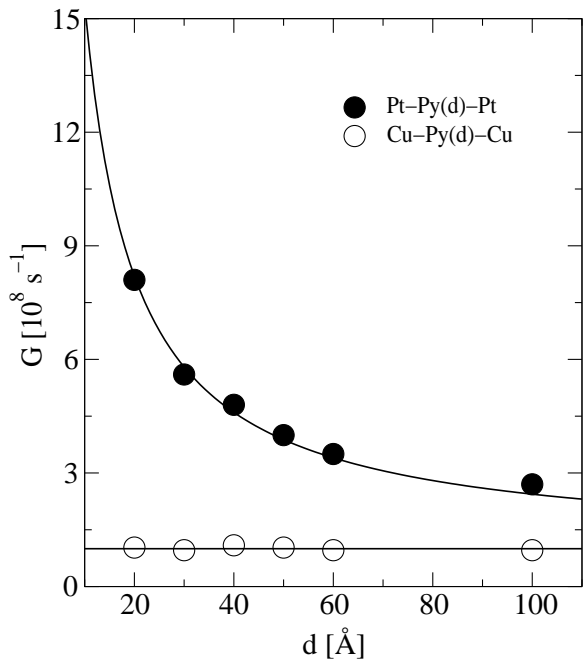}
\caption{Circles show measured (Ref.~\onlinecite{Mizukami:mmm01}) Gilbert
parameter G of a permalloy film with thickness $d$ sandwiched between two
normal-metal (Pt or Cu) layers. Solid lines are predictions of our theory
with two fitting parameters, $G_{0}$ and $g^{\uparrow \downarrow }$--Py
bulk damping and Py-Pt mixing conductance, respectively, see Eq.~(\ref{gd}).}
\label{f3}
\end{figure}

Infinite vs vanishing spin-flip rates in the normal metal are two
extreme regimes for the magnetization dynamics in \textit{F-N} bilayers. In
the former case, the damping constant $\alpha =\alpha
_{0}+g_{L}g^{\uparrow \downarrow }/(4\pi\mu)$ is significantly enhanced for
thin ferromagnetic films, whereas in the latter case, $\alpha =\alpha _{0}$
is independent of the ferromagnetic film thickness. Experimentally, the two
regimes are accessible by using Pt as a perfect or Cu as a poor spin sink in
contact with a ferromagnetic thin film, as done in
Ref.~\onlinecite{Mizukami:mmm01} for \textit{N}-Py-\textit{N} sandwiches.
(Using the \textit{N-F-N} trilayer simply increases $\alpha ^{\prime }$ by the
factor of 2, as compared to the \textit{F-N} bilayer, due to the spin
pumping through the two interfaces.) The measured damping parameter $
G=\gamma M_{s}\alpha $ is shown in Fig.~\ref{f3} by circles. 

For the Cu-Py-Cu trilayer, our theory predicts $G(d)=G_{0}$, while for the
Pt-Py-Pt sandwich 
\begin{equation}
G(d)=G_{0}+\frac{(g_{L}\mu _{\text{B}})^{2}}{2\pi \hbar }\frac{
g^{\uparrow \downarrow }S^{-1}}{d}  \label{gd}
\end{equation}
as a function of ferromagnetic film thickness $d$. The Py $g$ factor
is $g_{L}\approx 2.1$.\cite{Mizukami:mmm01} These expression agree with
the experiments for $G_{0}=1.0\times 10^{8}$~s$^{-1}$ and $g^{\uparrow
\downarrow }S^{-1}=2.6\times 10^{15}$~cm$^{-2}$ (see Fig.~\ref{f3}).
Both numbers are very reasonable: $G_{0}$ equals the bulk value $
0.7-1.0\times 10^{8}$~s$^{-1}$ for Py,\cite{Patton:jap75}
while $g^{\uparrow \downarrow }S^{-1}$ compares well with $
g^{\uparrow \downarrow }S^{-1}\approx 1.6\times 10^{15}$~cm$^{-2}$
found in angular-magnetoresistance (aMR) measurements in Py-Cu hybrids.\cite
{Bauer:prep}
(We recall that here one has to use the renormalized mixing
conductance $\tilde{g}^{\uparrow\downarrow}$, in the notation of
Ref.~\onlinecite{Bauer:prep}.)
In fact, since Pt has two conduction electrons per atom, while
Cu--only one, and they have similar crystal structures,
we expect $g^{\uparrow \downarrow }$ to be
larger in the case of the Py-Pt hybrid,
justifying the value used to fit the experimental data.
We have thus demonstrated that the additional damping in ferromagnetic
thin films can be used to measure the mixing conductance
of the \textit{F-N} interface.

\section{Magnetic damping in \textit{F-N1-N2} trilayer}

\label{fnn}

\begin{figure}[pth]
\includegraphics[scale=0.35,clip=]{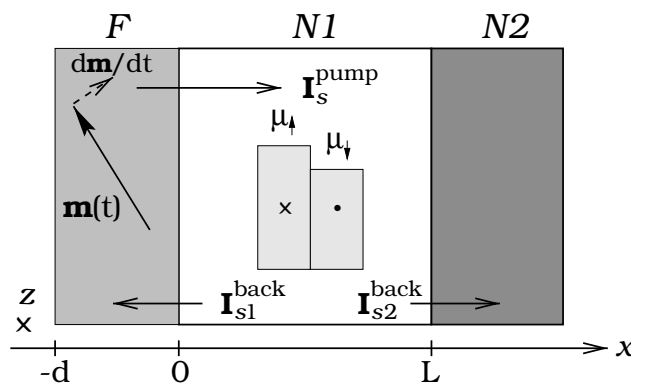}
\caption{Same as Fig.~\protect\ref{f2}, but now the normal-metal system is
composed of a bilayer \textit{N1-N2}. Ferromagnetic precession pumps spins
into the first normal-metal layer \textit{N1}. The spin build-up in
\textit{N1} may flow back into the ferromagnet \textit{F} as spin current
$\mathbf{I}_{s1}^{\text{back}}$, relax in \textit{N1}, or return to the second
normal-metal layer \textit{N2} as spin current $\mathbf{I}_{s2}^{\text{back}}
$. The spin accumulation in \textit{N2} is disregarded since the layer is
assumed to be a perfect spin sink.}
\label{f4}
\end{figure}

In this section we consider ferromagnetic spin pumping into a bilayer 
\textit{N1-N2} normal-metal system, see Fig.~\ref{f4}. It is assumed that
the spins are driven into the first normal-metal film (\textit{N1}) of
thickness $L$. While in \textit{N1}, spins are allowed to diffuse through
the film, where they can relax, diffuse back into the ferromagnet, or reach
the second normal-metal layer (\textit{N2}). \textit{N2} is taken to be a
perfect spin sink: spins reaching \textit{N2} either relax immediately by
spin-flip processes or are carried away before diffusing back into
\textit{N1}. We show that measuring the ferromagnetic magnetization damping
as a function of $L$ in this configuration can be used to study the dc mixing
conductance of the two \textit{N1} film interfaces as well as the \textit{N1}
spin-diffusion time.

\begin{figure}[pth]
\includegraphics[scale=0.45,clip=]{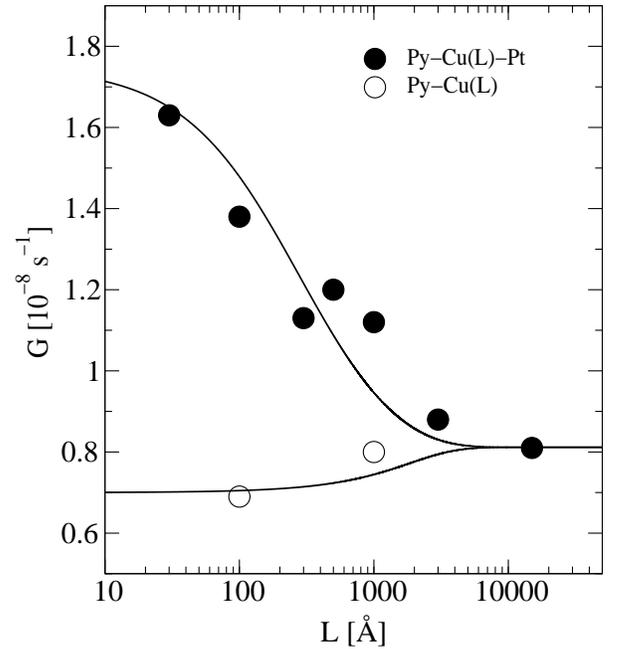}
\caption{Circles show the measurements by Mizukami \textit{et al.}
(Ref.~\onlinecite{Mizukami:mmm02}) of the Gilbert damping in Py-Cu-Pt trilayer
and Py-Cu bilayer as a function of the Cu buffer thickness $L$. Solid lines
are our theoretical prediction according to Eqs.~(\ref{g1}) and (\ref{g2}).}
\label{f5}
\end{figure}

The analysis in this section was inspired by experiments of Mizukami
\textit{et al.},\cite{Mizukami:mmm02} who in a follow-up to their systematic
study of Gilbert damping in \textit{N}-Py-\textit{N} sandwiches,\cite
{Mizukami:mmm01} studied magnetization damping in Py-Cu and Py-Cu-Pt hybrids as
a function of Cu film thickness $L$. The measured damping parameter $G$ is
shown by circles in Fig.~\ref{f5}. As shown in the preceding section, Cu is
a poor sink for the pumped spins, while Pt is nearly a perfect spin
absorber, thus identifying the Cu film with \textit{N1} and the Pt layer
with \textit{N2}.

We use the same notation as in the previous section to discuss the
\textit{F-N1} spin pumping with subsequent spin diffusion through
\textit{N1}. Similar to Eqs.~(\ref{bc}), the boundary conditions for the
diffusion equation (\ref{de}) in the normal metal \textit{N1} are now: 
\begin{align}
x=0:~\partial_{x}\text{\boldmath$\mu$}_{s} & =-2(\hbar\mathcal{N}SD)^{-1} 
\mathbf{I}_{s1}\,,  \notag \\
x=L:~\partial_{x}\text{\boldmath$\mu$}_{s} & =-2(\hbar\mathcal{N}SD)^{-1} 
\mathbf{I}_{s2}\,.  \label{bc1}
\end{align}
$\mathbf{I}_{s1}$ and $\mathbf{I}_{s2}$ are the total spin currents through
the left ($x=0$) and right ($x=L$) interfaces, respectively. $\mathbf{I}_{s1}
$ (similarly to $\mathbf{I}_{s}$ [Eq.~(\ref{sc})] in the previous
section) includes the pumped spin current [Eq.~(\ref{Is})] and the
spin--accumulation-driven spin current [Eq.~(\ref{ne})] contributions. $
\mathbf{I}_{s2}$, on the other hand, is entirely governed by the 
\textit{N1$\rightarrow$N2} spin--accumulation-driven flow 
\begin{equation}
\mathbf{I}_{s2}=\frac{g}{4\pi}\text{\boldmath$\mu$}
_{s}(x=L)\,,
\end{equation}
where $g$ is the conductance per spin of the \textit{N1-N2}
interface.

Solving the diffusion equation (\ref{de}) with the boundary conditions (\ref
{bc1}), we find the spin current $\mathbf{I}_{s1}$ as we did in the
preceding section. The
Gilbert damping enhancement due to the spin
relaxation in the composite normal-metal system is then given by
\begin{equation}
\alpha^{\prime}=\left[ 1+g^{\uparrow\downarrow}\frac{\tau_{\text{sf}
}\delta_{\text{sd}}}{h}\frac{1+\tanh(L/\lambda_{\text{sd}})g\tau_{\text{sf}
}\delta_{\text{sd}}/h}{\tanh(L/\lambda_{\text{sd}})+g\tau_{\text{sf}}
\delta_{\text{sd}}/h}\right] ^{-1}\frac{g_{L}g^{\uparrow\downarrow}}
{4\pi\mu}\,.
\label{ab}
\end{equation}
Setting $g=0$ decouples the two normal-metal systems and
reduces Eq.~(\ref{ab}) to Eq.~(\ref{aa})
giving the damping coefficient of the \textit{F-N1}
bilayer.
From Eq.~(\ref{ab}), we get for the Py-Cu$(L)$-Pt trilayer 
\begin{align}
G(L)& =G_{0}+\left[ 1+g^{\uparrow \downarrow }\frac{\tau_{\text{sf}}
\delta_{\text{sd}}}{h}\frac{1+\tanh (L/\lambda_{\text{sd}})g\tau_{\text{sf}}
\delta_{\text{sd}}/h}{\tanh(L/\lambda_{\text{sd}})+g\tau_{\text{sf}}
\delta_{\text{sd}}/h}\right] ^{-1}  \notag \\
& \times \frac{(g_{L}\mu_{\text{B}})^{2}}{2h}\frac{g^{\uparrow
\downarrow }S^{-1}}{d}  \label{g1}
\end{align}
and for the Py-Cu$(L)$ bilayer (putting $g=0$)
\begin{equation}
G(L)=G_{0}+\left[ 1+\frac{g^{\uparrow \downarrow }\tau_{\text{sf}}
\delta_{\text{sd}}/h}{\tanh(L/\lambda_{\text{sd}})}\right]^{-1}\frac{
(g_{L}\mu_{\text{B}})^{2}}{2h}\frac{g^{\uparrow \downarrow }S^{-1}}{d}
\,.  \label{g2}
\end{equation}
In the experiments, the permalloy thickness $d=30$~\AA\ is fixed and the Cu
film thickness $L$ is varied between 3 and 1500 nm\ as shown by the circles
in Fig.~\ref{f5}. Our theoretical results, Eqs.~(\ref{g1}) and (\ref{g2}),
are plotted in Fig.~\ref{f5} by solid lines. We use the following
parameters: The bulk damping\cite{Patton:jap75}
$G_{0}=0.7\times 10^{8}$~s$^{-1}$; the spin-flip probability $\epsilon
=1/700$ and the spin-diffusion length $\lambda _{\text{sd}}=250$~nm for
Cu (which correspond to elastic mean free path
$\lambda_{\text{el}}=\sqrt{3\epsilon}\lambda _{\text{sd}}=16$~nm),
in agreement with values reported in literature;\cite
{Meservey:prl78,Yang:prl94,Jedema:nat01} $g^{\uparrow \downarrow
}S^{-1}=1.6\times 10^{15}$~cm$^{-2}$ from the
aMR measurements;\cite{Bauer:prep} and
$gS^{-1}=3.5\times 10^{15}$~cm$^{-2}$ for the Cu-Pt contact,
which lies between values for the majority and minority carriers
as measured and calculated\cite{Xia:prb01} for the Cu-Co interface.
Figure~\ref{f5} shows a remarkable
agreement (within the experimental error) between the measurements and
our theory. It is important to stress that while the profiles of the trends
displayed in Fig.~\ref{f5} reveal the diffusive nature of spin transfer
in the Cu spacer, they cannot be used to judge the validity of a detailed
mechanism for spin injection (relaxation) at the Py-Cu (Cu-Pt) interface.
The case of our spin pumping picture is strongly supported by the
normalization of the curves (in agreement with experiment), which are
governed in our theory by quantities known from other sources. 

The trends in Fig.~\ref{f5} can be understood as follows. Since Cu is a
poor spin sink, a Py-Cu contact with a single Cu film does not lead to a
significant damping enhancement. The small spin-flip ratio, $\epsilon\ll1$,
causes most of the spins transferred into the normal-metal layer to be
scattered back and relax in the ferromagnet before flipping their direction
in the Cu buffer. This leads only to a small damping enhancement, which
saturates at $L\gg\lambda_{\text{sd}}$ and vanishes in the limit $
L\ll\lambda _{\text{sd}}$. The situation changes after a Pt film, a very
good spin sink, is connected to the bilayer: If the normal-metal layer is
smaller than the elastic mean free path, $L\ll\lambda_{\text{el}}$, the spin
accumulation is uniform throughout the Cu buffer. The spin pumping will now
be partitioned. A fraction of the pumped spins reflects back into the
ferromagnet, while the rest get transmitted and subsequently relax in the Pt
layer. The ratio between these two fractions equals the ratio between the
conductance of the Py-Cu contact and the Co-Pt contact, $g
^{\uparrow\downarrow}/g$, and is of the order of unity. This
results in a large magnetization damping as a significant portion of the
spin pumping relaxes by spin-orbit scattering in Pt. When $L$ is increased,
less spins manage to diffuse through the entire Cu buffer, and, in the limit 
$L\gg\lambda_{\text{sd}}$, the majority of the spins scatter back into the
ferromagnet or relax in Cu not feeling the presence of the Pt layer at all.
In the intermediate regime, the spin pumping into the Pt layer has
an algebraic fall-off on the scale of the elastic mean free path and
exponential one on the scale of the spin-diffusion length.

It is important to emphasize that the strong dependence of
damping on the Cu layer thickness $L$ in the Py-Cu-Pt configuration gives
evidence of the spin accumulation in the normal-metal system. This spin
accumulation, in turn, indicates that an excited ferromagnet (as in the FMR
experiment discussed here) transfers spins into adjacent nonmagnetic layers,
confirming our claim.\cite{Tserkovnyak:prl02} Furthermore, this supports
our concept of the spin battery.\cite{Brataas:prb02}

Before ending this section, it is illuminating to make a small
digression and further study Eq.~(\ref{g1}) in the limit of
vanishing spin-flip processes in the buffer layer \textit{N1}.
Recalling our definitions for $\lambda_{\text{sd}}$ and $\delta_{\text{sd}}$
[Eqs.~(\ref{sd}) and (\ref{td})] and taking limit
$\tau_{\text{sf}}\rightarrow\infty$, we find that Eq.~(\ref{g1}) reduces to
Eq.~(\ref{gd}), only with $g^{\uparrow\downarrow}$ replaced by
$g_{\text{eff}}^{\uparrow\downarrow}$ [similarly to Eq.~(\ref{ser})]:
\begin{equation}
\frac{1}{g_{\text{eff}}^{\uparrow\downarrow}}=
\frac{1}{g^{\uparrow\downarrow}}+R_{N1}+
\frac{1}{g}\,,
\label{geff}
\end{equation}
where $R_{N1}$ is the resistance of the \textit{N1} layer.
The right-hand side of Eq.~(\ref{geff}) is simply the inverse
mixing conductance of
the \textit{N1} buffer in series with its two interfaces (one with \textit{F}
and one with \textit{N2});\cite{Bauer:prep} in particular,
when layer \textit{N1} is thick enough, the total mixing
conductance $g_{\text{eff}}^{\uparrow\downarrow}$
is just the conductance of the diffusive normal-metal spacer
separating \textit{F} and \textit{N2}.\cite{Brataas:prl00,Brataas:epjb01}
The spin pumping into
layer \textit{N1} with the subsequent diffusion and then spin absorption
by the ideal spin sink \textit{N2} (as discussed in this section)
can thus be viewed as the spin pumping across an effective combined
scatterer separating the ferromagnet (\textit{F}) from the perfect spin sink
(\textit{N2}) [as done in obtaining Eq.~(\ref{gd})].
This shows consistency of our approach.

\section{Conclusions and discussions}

\label{cd}

Ferromagnets emit a spin current into adjacent normal metals when the
magnetization direction changes with time. We recently proposed a novel
mechanism for this spin transfer based on the picture of adiabatic spin
pumping.\cite{Tserkovnyak:prl02} It was shown that our theory explains
the increased magnetization damping in ferromagnets in contact with normal
metals in measurements of FMR linewidths.\cite
{Mizukami:mmm01,Mizukami:mmm02,Heinrich:prl87,Urban:prl01}

Whereas the spin pumping affects the magnetization dynamics, it also creates
a nonequilibrium magnetization in adjacent nonmagnetic films. In this paper
we first calculate this spin accumulation for \textit{F-N} metallic
multilayers and find that it induces a spin backflow into the
ferromagnetic layer that reduces the spin pumping. This
spin--accumulation-driven current is significant for light metals or metals
with only \textit{s} electrons in the conduction band, which have a small
spin-flip to spin-conserving scattering ratio.

The picture of ferromagnetic spin pumping and subsequent
spin diffusion in the adjacent normal-metal layers is also applied to the
\textit{F-N1-N2} configuration in order to analyze recent
experiments\cite{Mizukami:mmm02} on magnetization damping in Py-Cu-Pt
trilayers. We showed that our theory quantitatively explains the experimental
findings. Our analysis of the experiments by Mizukami \textit{et al.} \cite
{Mizukami:mmm01,Mizukami:mmm02} shows that FMR of ultrathin
ferromagnetic films in contact with single or composite normal-metal buffers
is a powerful tool to investigate interfacial transport properties of magnetic
multilayers as well as the spin relaxation parameters of the normal-metal
layers.

\acknowledgments

We are grateful to B.~I.~Halperin and Yu.~V.~Nazarov for stimulating
discussions. This work was supported in part by the DARPA Award No. MDA
972-01-1-0024, the NEDO International Joint Research Grant Program
\textquotedblleft Nano-magnetoelectronics\textquotedblright, NSF Grant DMR
99-81283, the FOM, and the Schlumberger Foundation.

\appendix

\section*{Appendix: Adiabatic spin pumping}

\begin{figure}[pth]
\includegraphics[scale=0.45]{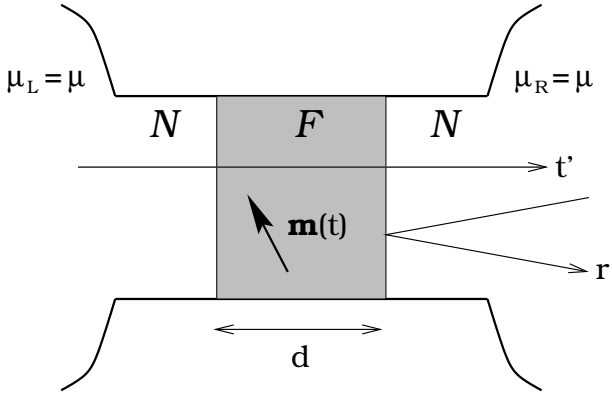}
\caption{Ferromagnetic film (\textit{F}) sandwiched between two normal-metal
layers (\textit{N}). The latter are taken to be reservoirs in common thermal
equilibrium. The reflection and transmission amplitudes $r$ and $t^{\prime }$
shown here govern the spin current pumped into the right lead.}
\label{f6}
\end{figure}

Here we present a detailed discussion of spin pumping into normal-metal
layers by a precessing magnetization direction $\mathbf{m}$ of an adjacent
ferromagnet. A schematic of the model is displayed in Fig.~\ref{f6}. The
ferromagnetic layer \textit{F} is a a spin-dependent scatterer that governs
electron transport between two [left ($L$) and right ($R$)] normal-metal
reservoirs.

The $2\times 2$ operator $\hat{I}_{l}$ for the charge and spin current in
the $l$th lead $\left( l=L,R\right) $ can be expressed in terms of
operators $a_{\alpha m,l}(E)$ [$b_{\alpha m,l}(E)$] that annihilate a spin-$
\alpha $ electron with energy $E$ leaving [entering] the $l$th lead through
the $m$th channel: 
\begin{align}
\hat{I}_{l}^{\alpha \beta }(t)& =\frac{e}{h}\sum_{m}\int dEdE^{\prime
}e^{i(E-E^{\prime })t/\hbar }  \notag  \label{I} \\
& \times \left[ a_{\beta m,l}^{\dagger }(E)a_{\alpha m,l}(E^{\prime
})-b_{\beta m,l}^{\dagger }(E)b_{\alpha m,l}(E^{\prime })\right] \,.
\end{align}
When the scattering matrix $\hat{s}_{mn,ll^{\prime }}^{\alpha \beta }(t)$ of
the ferromagnetic layer varies slowly on the time scales of electronic
relaxation in the system, an adiabatic approximation may be used. The
annihilation operators for particles entering the reservoirs are then
related to the operators of the outgoing states by the instantaneous value
of the scattering matrix: $b_{\alpha m,l}(E)=\hat{s}_{mn,ll^{\prime
}}^{\alpha \beta }(t)a_{\beta n,l^{\prime }}(E)$. In terms of $a_{\alpha m,l}
$ only, we can evaluate the expectation value $\left\langle \hat{I}
_{l}^{\alpha \beta }(t)\right\rangle $ of the current operator using $
\langle a_{\alpha m,l}^{\dagger }(E)a_{\beta n,l^{\prime }}(E^{\prime
})\rangle =f_{l}(E)\delta _{\alpha \beta }\delta _{mn}\delta _{ll^{\prime
}}\delta (E-E^{\prime })$, where $f_{l}(E)$ is the (isotropic) distribution
function in the $l$th reservoir. When the scattering matrix depends on a
single time-dependent parameter $X(t)$, then the Fourier transform of the
current expectation value $\hat{I}_{l}(\omega )=\int dte^{i\omega t}\hat{I}
_{l}(t)$ can be written as 
\begin{equation}
\hat{I}_{l}(\omega )=\hat{g}_{X,l}(\omega )X(\omega )  \label{Il}
\end{equation}
in terms of a frequency $\omega $- and $X$-dependent parameter $\hat{g}_{X,l}
$:\cite{Buttiker:zpb94} 
\begin{align}
\hat{g}_{X,l}(\omega )& =-\frac{e\omega }{4\pi }\sum_{l^{\prime }}\int
dE\left( -\frac{\partial f_{l^{\prime }}(E)}{\partial E}\right)   \notag \\
& \times \sum_{mn}\left( \frac{\partial \hat{s}_{mn,ll^{\prime }}(E)}{
\partial X}\hat{s}_{mn,ll^{\prime }}^{\dagger }(E)-\text{H.c.}\right) \,.
\end{align}
Equation~(\ref{Il}) is the first-order (in frequency) correction to the dc
Landauer-B{\"{u}}ttiker formula.\cite{Buttiker:prl86} At equilibrium $
f_{R}(E)=f_{L}(E)$, Eq.~(\ref{Il}) is the lowest-order nonvanishing
contribution to the current. Furthermore, at sufficiently low temperatures,
we can approximate $-\partial f_{l}(E)/\partial E$ by a $\delta $-function
centered at Fermi energy. The expectation value of the $2\times 2$
particle-number operator $\hat{Q}_{l}(\omega )$ [defined by $\hat{I}_{l}(t)=d
\hat{Q}_{l}(t)/dt$ in time or by $\hat{I}_{l}(\omega )=-i\omega \hat{Q}
_{l}(\omega )$ in frequency domain] for the $l$th reservoir is then given
by 
\begin{equation}
\hat{Q}_{l}(\omega )=\left( \frac{e}{4\pi i}\sum_{mnl^{\prime }}\frac{
\partial \hat{s}_{mn,ll^{\prime }}}{\partial X}\hat{s}_{mn,ll^{\prime
}}^{\dagger }+\text{H.c.}\right) X(\omega )\,,  \label{Q}
\end{equation}
where the scattering matrices are evaluated at the Fermi energy. Because the
prefactor on the right-hand side of Eq.~(\ref{Q}) does not depend on
frequency $\omega $, the equation is also valid in time domain. The change
in particle number $\delta \hat{Q}_{l}(t)$ is proportional to the modulation 
$\delta X(t)$ of parameter $X$ and the $2\times 2$ matrix current (directed
into the normal-metal leads) reads 
\begin{equation}
\hat{I}_{l}(t)=e\frac{\partial \hat{n}_{l}}{\partial X}\frac{dX(t)}{dt}\,,
\label{Ip}
\end{equation}
where the \textquotedblleft matrix emissivity\textquotedblright\ into lead $l
$ is 
\begin{equation}
\frac{\partial \hat{n}_{l}}{\partial X}=\frac{1}{4\pi i}\sum_{mnl^{\prime }}
\frac{\partial \hat{s}_{mn,ll^{\prime }}}{\partial X}\hat{s}_{mn,ll^{\prime
}}^{\dagger }+\text{H.c.}\,.  \label{em}
\end{equation}
If the spin-flip scattering in the ferromagnetic layer is disregarded, the
scattering matrix $\hat{s}$ can be written in terms of the spin-up and
spin-down scattering coefficients $s^{\uparrow (\downarrow )}$ using the
projection matrices $\hat{u}^{\uparrow }=\left( \hat{1}+\hat{\text{\boldmath$
\sigma $}}\cdot \mathbf{m}\right) /2$ and $\hat{u}^{\downarrow }=\left( \hat{
1}-\hat{\text{\boldmath$\sigma $}}\cdot \mathbf{m}\right) /2$:\cite
{Brataas:prl00} 
\begin{equation}
\hat{s}_{mn,ll^{\prime }}=s_{mn,ll^{\prime }}^{\uparrow }\hat{u}^{\uparrow
}+s_{mn,ll^{\prime }}^{\downarrow }\hat{u}^{\downarrow }\,.  \label{s}
\end{equation}
The spin current pumped by the magnetization precession is obtained by
identifying $X(t)=\varphi (t)$, where $\varphi $ is the azimuthal angle of
the magnetization direction in the plane perpendicular to the precession
axis. For simplicity, we assume that the magnetization \textit{rotates}
around the $y$ axis: $\mathbf{m}=(\sin \varphi ,0,\cos \varphi )$. Using
Eq.~(\ref{s}), it is then easy to calculate the emissivity [Eq.~(\ref{em})]
for this process: 
\begin{equation}
\frac{\partial \hat{n}_{l}}{\partial \varphi }=-\frac{1}{4\pi }\left[
A_{r}\sigma _{y}+A_{i}(\sigma _{x}\cos \varphi -\sigma _{z}\sin \varphi )
\right] \,,  \label{nl}
\end{equation}
where $A_{r}(A_{i})=\mbox{Re}(\mbox{Im})[g^{\uparrow \downarrow
}-t^{\uparrow \downarrow }]$, as explained in Sec.~\ref{pump}. Expanding the 
$2\times 2$ current into isotropic and traceless components, 
\begin{equation}
\hat{I}=\frac{\hat{1}}{2}I_{c}-\frac{e}{\hbar }\hat{\text{\boldmath$\sigma $}
}\cdot \mathbf{I}_{s}\,,  \label{Ii}
\end{equation}
we identify the charge current $I_{c}$ and spin current $\mathbf{I}_{s}$.
Comparing Eqs.~(\ref{Ip}), (\ref{nl}), and (\ref{Ii}), we find that the
charge current vanishes, $I_{c}=0$, and the spin current 
\begin{equation}
\mathbf{I}_{s}=\left( A_{i}\cos \varphi ,A_{r},-A_{i}\sin \varphi \right) 
\frac{\hbar }{4\pi }\frac{d\varphi }{dt}
\end{equation}
can be rewritten as Eq.~(\ref{Is}). Because the spin current transforms as a
vector, it is straightforward to show that Eq.~(\ref{Is}) is also valid in
the case of the general motion of the magnetization direction.

Even though the mathematics of our scattering approach to adiabatic spin
pumping is entirely analogous to the charge-pumping theory developed in
Ref.~\onlinecite{Brouwer:prb98}, there are some striking differences in the
physics. In the case of a spin-independent scatterer as in
Ref.~\onlinecite{Brouwer:prb98},
the average charge-pumping current has the same
direction in the two leads, by charge conservation: the charge entering the
scattering region through either lead must leave it within a period of the
external-gate variations. Whereas the particle number of the two reservoirs
must (on average) be conserved also here, the total conduction-electron spin
angular momentum is not conserved. In fact, as we explained in
Ref.~\onlinecite{Tserkovnyak:prl02} for a symmetric
system shown in Fig.~\ref{f6},
a precessing ferromagnet loses angular momentum by polarizing adjacent
nonmagnetic conductors. In this respect, the phenomenon looks more similar to a
spin \textquotedblleft well\textquotedblright\ or \textquotedblleft
fountain\textquotedblright : An excited ferromagnet ejects spins in all
directions into adjacent conductors by losing its own angular momentum,
rather than transfers (\textquotedblleft pumps\textquotedblright ) spins
from one lead to the other. The angular momentum has to be provided, of
course, by the applied magnetic field.

\end{document}